\documentclass[aps,prl,showpacs,twocolumn,superscriptaddress]{revtex4}
\usepackage{multirow}
\usepackage{graphicx}
\usepackage{amsmath}
\newcommand{\sqrtsNN}{\mbox{$\sqrt{s_{_{NN}}}$}}
\newcommand{\axi}{$\overline{\Xi}^+$}
\newcommand{\xim}{$\Xi^-$}
\newcommand{\omegam}{$\Omega^-$}
\newcommand{\aomega}{$\overline{\Omega}^+$}
\begin{document}
\title{Multi-Strange Baryon Production in Au-Au collisions at
\sqrtsNN\ =~130~GeV}
%

\affiliation{Argonne National Laboratory, Argonne, Illinois 60439}
\affiliation{Brookhaven National Laboratory, Upton, New York 11973}
\affiliation{University of Birmingham, Birmingham, United Kingdom}
\affiliation{University of California, Berkeley, California 94720}
\affiliation{University of California, Davis, California 95616}
\affiliation{University of California, Los Angeles, California 90095}
\affiliation{Carnegie Mellon University, Pittsburgh, Pennsylvania 15213}
\affiliation{Creighton University, Omaha, Nebraska 68178}
\affiliation{Nuclear Physics Institute AS CR, \v{R}e\v{z}/Prague, Czech Republic}
\affiliation{Laboratory for High Energy (JINR), Dubna, Russia}
\affiliation{Particle Physics Laboratory (JINR), Dubna, Russia}
\affiliation{University of Frankfurt, Frankfurt, Germany}
\affiliation{Indiana University, Bloomington, Indiana 47408}
\affiliation{Insitute  of Physics, Bhubaneswar 751005, India}
\affiliation{Institut de Recherches Subatomiques, Strasbourg, France}
\affiliation{University of Jammu, Jammu 180001, India}
\affiliation{Kent State University, Kent, Ohio 44242}
\affiliation{Lawrence Berkeley National Laboratory, Berkeley, California 94720}\affiliation{Max-Planck-Institut f\"ur Physik, Munich, Germany}
\affiliation{Michigan State University, East Lansing, Michigan 48824}
\affiliation{Moscow Engineering Physics Institute, Moscow Russia}
\affiliation{City College of New York, New York City, New York 10031}
\affiliation{NIKHEF, Amsterdam, The Netherlands}
\affiliation{Ohio State University, Columbus, Ohio 43210}
\affiliation{Panjab University, Chandigarh 160014, India}
\affiliation{Pennsylvania State University, University Park, Pennsylvania 16802}
\affiliation{Institute of High Energy Physics, Protvino, Russia}
\affiliation{Purdue University, West Lafayette, Indiana 47907}
\affiliation{University of Rajasthan, Jaipur 302004, India}
\affiliation{Rice University, Houston, Texas 77251}
\affiliation{Universidade de Sao Paulo, Sao Paulo, Brazil}
\affiliation{University of Science \& Technology of China, Anhui 230027, China}
\affiliation{Shanghai Institute of Nuclear Research, Shanghai 201800, P.R. China}
\affiliation{SUBATECH, Nantes, France}
\affiliation{Texas A\&M, College Station, Texas 77843}
\affiliation{University of Texas, Austin, Texas 78712}
\affiliation{Valparaiso University, Valparaiso, Indiana 46383}
\affiliation{Variable Energy Cyclotron Centre, Kolkata 700064, India}
\affiliation{Warsaw University of Technology, Warsaw, Poland}
\affiliation{University of Washington, Seattle, Washington 98195}
\affiliation{Wayne State University, Detroit, Michigan 48201}
\affiliation{Institute of Particle Physics, CCNU (HZNU), Wuhan, 430079 China}
\affiliation{Yale University, New Haven, Connecticut 06520}
\affiliation{University of Zagreb, Zagreb, HR-10002, Croatia}
\author{J.~Adams}\affiliation{University of Birmingham, Birmingham, United Kingdom}
\author{C.~Adler}\affiliation{University of Frankfurt, Frankfurt, Germany}
\author{M.M.~Aggarwal}\affiliation{Panjab University, Chandigarh 160014, India}
\author{Z.~Ahammed}\affiliation{Purdue University, West Lafayette, Indiana 47907}
\author{J.~Amonett}\affiliation{Kent State University, Kent, Ohio 44242}
\author{B.D.~Anderson}\affiliation{Kent State University, Kent, Ohio 44242}
\author{M.~Anderson}\affiliation{University of California, Davis, California 95616}
\author{D.~Arkhipkin}\affiliation{Particle Physics Laboratory (JINR), Dubna, Russia}
\author{G.S.~Averichev}\affiliation{Laboratory for High Energy (JINR), Dubna, Russia}
\author{S.K.~Badyal}\affiliation{University of Jammu, Jammu 180001, India}
% ...

\collaboration{STAR Collaboration}
\homepage{www.star.bnl.gov}\noaffiliation


\date{\today}

\begin{abstract}
 The transverse mass spectra and mid-rapidity yields for
$\Xi$s and $\Omega$s plus their anti-particles are presented.
 The 10$\%$ most central collision yields suggest that the
amount  of multi-strange particles produced per produced
charged hadron  increases from SPS to RHIC energies.
 A hydrodynamically inspired model fit to the spectra, which
assumes a thermalized source, seems to indicate that these
multi-strange particles experience a significant transverse
flow effect, but are emitted when the system is hotter and the
flow is smaller than values obtained from a combined fit to
$\pi$, K, p and $\Lambda$s.
\end{abstract}

\maketitle

In heavy ion collisions we aim to investigate nuclear matter under
extreme conditions of pressure and temperature which is expected
to lead to the creation of deconfined partonic matter, the Quark
Gluon Plasma (QGP)~\cite{QM02}.
The study of strange particles production is thought to yield
information on the collision dynamics from the early stage to
chemical and thermal freeze-out. The production of strangeness
through partonic interactions, mainly $gg \rightarrow
s\overline{s}$, is expected to dominate over that by hadronic
scatterings~\cite{MR82,KMR86}. Since at RHIC energies the initial
gluon density is expected to be much higher than in lower-energy
heavy ion collisions~\cite{PartonDensity1,PartonDensity2}, strange
particles, especially multi-strange particles are expected to be
more sensitive to the partonic dynamics of the early stage of the
collision.

Chemical freeze-out is defined by the temperature,
$T\mathrm{_{ch}}$, at which inelastic collisions cease and the
relative particle ratios become fixed.
 With no strange quarks in the initial state it is possible for
the system to freeze-out chemically before absolute strangeness
chemical equilibrium has been reached.
 Statistical models have had much success in reproducing the
particle ratios both at lower energies
(e.g.~\cite{ThermMod1,ThermMod2,ThermMod3}) and at RHIC
(e.g.~\cite{ThermMod4,ThermMod5}).
 Some of these  models, which assume chemical equilibrium,
allow for the possibility of non-equilibrium in the strangeness
channel by the introduction of a strangeness saturation factor,
$\gamma_{s}$~\cite{ThermMod2,ThermMod5}.
 This factor reaches unity when complete strangeness saturation,
or absolute chemical equilibrium, is present in the system.
 A common $T\mathrm{_{ch}}$ of 170
MeV~\cite{ThermMod4} is indicated by fits to measurements at 130
GeV of anti-baryon to baryon ratios as well as particle ratios
involving only mesons and non-strange baryons.
This temperature is remarkably close to that of the phase
transition calculated with lattice QCD~\cite{Lqcd}.
Re-calculating the fits with the inclusion of the multi-strange
baryons allows a more sensitive test of statistical models to be
made.

A comprehensive analysis of the transverse momentum or transverse
mass spectra allows us to probe the possibility of thermal
equilibrium of the system.
 The high mass of the multi-strange baryons means they are more
sensitive to the size of the transverse flow of the system.
 It has previously been suggested that thermal freeze-out occurs
much earlier for $\Omega$ and $\Xi$ than for lower mass particles
due to the predicted low scattering cross section of $\Omega$ and
$\Xi$~\cite{OmXSec}.
 In the limit of vanishing cross sections this would mean that
these particles are emitted almost directly from the phase
boundary of the hadronizing fireball.
 However, these cross sections have not actually been measured and may not be
negligible. If they were significant, combined with the now
sizeable particle production at these energies, the $\Omega$ + X
elastic collision rate would delay the $\Omega$ kinetic freeze-out,
resulting in these particles carrying the same flow velocity as
the pions and kaons.
 By comparing different thermal freeze-out scenarios to the
data (e.g.~\cite{Bugaev, Broniowski, Teaney}) it may be possible
to determine the rate of build-up of radial flow.

\begin{figure}
\begin{center}
\includegraphics[width=0.48\textwidth]{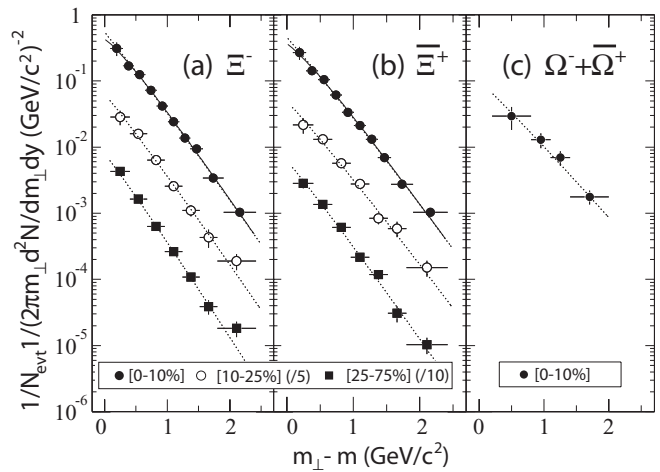}
\caption{ $m_{\bot}$ spectra for (a) \xim, (b) \axi and (c)
$\Omega$ as a function of centrality.
Scale factors  have been applied to the spectra for clarity.
Points are drawn at the bin center.
The horizontal bars indicate the bin size.
The dashed curves are Boltzmann fits to the spectra.
The solid curves are hydrodynamically inspired model fits to
the most central \xim and \axi spectra.}
\label{Fig:BoltMt}
\end{center}
\end{figure}

\begin{table}[!htb]
\begin{center}
\begin{ruledtabular}
\begin{tabular} {rlclcl}
\multicolumn{1}{c}{$h^{-}$}& &\multicolumn{2}{c}{Exponential}&\multicolumn{2}{c}{Boltzmann}\\
             &                &$dN/dy$     &\multicolumn{1}{c}{$T_{\mathrm{E}}$(MeV)}     & $dN/dy$ & \multicolumn{1}{c}{$T_{\mathrm{B}}$(MeV)}        \\
\hline
 \multirow{3}{1.8cm}{260.3$\pm$7.5}  &$\Xi^{-}$       &2.16$\pm$0.09 &338$\pm$6  &2.06$\pm$0.09 &296$\pm$5  \\
   & $\bar{\Xi}^{+}$ &1.81$\pm$0.08 &339$\pm$7  &1.73$\pm$0.08 &297$\pm$5  \\
   & $\Omega$        &0.59$\pm$0.14 &417$\pm$52 &0.58$\pm$0.14 &362$\pm$39 \\
\hline
\multirow{2}{1.8cm}{163.6$\pm$5.2} &$\Xi^{-}$  &1.22$\pm$0.11 &335$\pm$16 &1.18$\pm$0.11 &291$\pm$13 \\
  &$\bar{\Xi}^{+}$ &1.00$\pm$0.10 &349$\pm$17  &0.97$\pm$0.10 &302$\pm$13 \\
\hline
 \multirow{2}{1.8cm}{42.5$\pm$3.0} &$\Xi^{-}$       &0.28$\pm$0.02 &312$\pm$12 &0.27$\pm$0.02 &273$\pm$10 \\
  &$\bar{\Xi}^{+}$ &0.23$\pm$0.02 &320$\pm$11 &0.22$\pm$0.02 &280$\pm$9  \\
\end{tabular}
\end{ruledtabular}
\end{center}
\caption{ Fit parameters for the $m_{\bot}$ spectra of the \xim,
\axi and $\Omega$.
 The data represent the 0-10$\%$, 10-25$\%$ and 25-75$\%$
centrality bins with
$h^{-}$=$dN_{h^{-}}$/d$\eta$\big|$_{|\eta|<0.5}$.
 Only statistical and $p_{\bot}$ dependent systematic uncertainties
are presented.
 The $p_{\bot}$ independent systematic uncertanties are 10$\%$.}
\label{Table:MtFits}
\end{table}

\begin{figure*}[tb]
\begin{center}
 \includegraphics[width=0.63\textwidth]{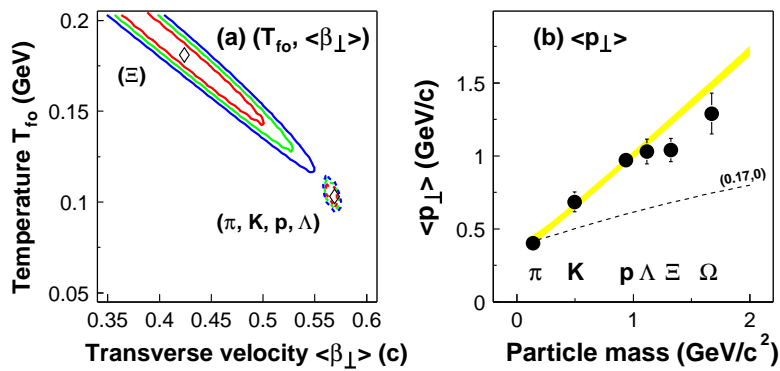}
 \caption{ (a) The kinetic freeze-out temperature vs transverse flow
velocity for the hydrodynamically inspired model fits to the
$m_{\bot}$ spectra.
 The 1, 2 and 3 sigma contours are shown.
 Solid curves are for a simultaneous fit to the \xim and \axi.
 Dashed curves are a separate fit to the STAR $\pi$, K, p and
$\Lambda$ data.
 The diamonds represent the best fit in both cases.
 (b) Mean transverse momenta for identified particles vs particle
mass (see text for details).
 The band results from the three sigma contour of the hydrodynamically
inspired model fit to the $\pi$, K, p and $\Lambda$ data and the dashed
curve is for $T_{\mathrm{fo}}=170$ MeV, $<\beta_{\bot}>=0$.}
\label{Fig:HydroCont}
\end{center}
\end{figure*}

 The data presented here were taken with the STAR detector in Au+Au
collisions at $\sqrt{s_{_{NN}}}=130$ GeV.
 The main tracking detector is a large cylindrical Time Projection Chamber.
 A Central Trigger Barrel measuring the produced charged particle
multiplicity around mid-rapidity plus two Zero Degree Calorimeters
essentially measuring  neutral spectator energy were used for
triggering~\cite{StarSetup}.
 In this paper the centrality of the collisions is determined from
the measured mid-rapidity negative particle multiplicity.
 The data were divided into three centrality
classes corresponding to 0-10$\%$, 10-25$\%$ and 25-75$\%$ of the
total hadronic cross section as described in~\cite{StarMultip}.
Multi-strange particles are identified via their decay modes $ \Xi
\rightarrow \Lambda + \pi$ and $\Omega \rightarrow \Lambda + K$
with the subsequent decay of $\Lambda \rightarrow p + \pi$.
 The tertiary $\Lambda$ vertex is identified by selecting positive
and negative tracks that are consistent with an origin at the
decay of a hyperon some distance from the primary collision
point~\cite{StarLambda}.
 The secondary vertex of the decay is located in a similar fashion
by combining the previously identified $\Lambda$ with a charged
particle.
 Simple cuts on geometry, kinematics and  particle identification,
via specific ionization, are applied at each step to reduce
the background due to the high multiplicity \cite{StarMultip}.
 The momenta of the daughter particles at the decay vertex are then
combined to calculate the parent particle kinematics.
 The peaks  in the invariant mass plots have an average signal to
noise ratio of $0.74$, $0.78$ and $0.86$ for the \xim, \axi and
$\Omega$ respectively in the 0-10$\%$ centrality bin.
 The signal to noise ratios are calculated for $\pm15$ MeV/c$^2$
about the expected mass.
 The statistics of the \omegam and \aomega signal are not sufficient
to allow a separate measurement of the spectra of each particle.
 Hence $\Omega$ refers to \omegam plus \aomega.
 The momentum integrated \aomega/\omegam ratio for the
top 11$\%$ most central data is
$0.95\pm0.15(stat)\pm0.05(sys)$~\cite{StarRatio}.

The invariant mass distributions are histogramed in transverse
mass, $m_{\bot}$=$\sqrt{p_{\bot}^{2} + m^{2}}$, and the signal
extracted for each bin.
 The raw yields are calculated from the invariant mass distributions
by counting the entries within $\pm15$ MeV/c$^2$ about the expected
mass and then subtracting the background.
 The background is estimated by sampling two regions on either side of
the peak.
 The raw yield in each $m_{\bot}$ bin is corrected for detector acceptance and
reconstruction efficiency by the Monte Carlo technique, where
simulated particles were embedded into real events.
 The data cover $\mid y \mid<0.75$, where efficiency and
acceptance studies have shown the corrections to be constant. The
total correction factors for the $\Xi$($\Omega$) are
$0.2\%$($0.04\%$) for the lowest $m_{\bot}$ bin rising to
$4.0\%$($0.5\%$) for the highest bin.

Fig.~\ref{Fig:BoltMt} shows the invariant $m_{\bot}$ spectra
as functions of centrality for the \xim and
\axi, and the $\Omega$ for the 10$\%$ most central data.
 A portion of the systematic uncertainties are
a function of $p_{\bot}$, therefore this uncertainty is added in
quadrature, on a bin-by-bin basis, to the statistical one,
yielding the vertical bars in Fig.~\ref{Fig:BoltMt}.
 We estimate the remaining systematic uncertainties to be $10\%$ on
both the extracted invariant yields and slope parameters, the major
source of which lies in the misrepresentation of the embedded
Monte-Carlo to the data resulting in a systematic uncertainty in
the efficiency calculation.
 These were obtained by exploring the dependence of the invariant
yields and slope parameters to changes in the cuts phase space
(more details can be found in~\cite{CurtisThes, JavThes,
BorisThes}).
 The weak decay feed-down  of $\Omega$ on $\Xi$ is estimated to be
less than $2\%$ and thus is neglected.

Table~\ref{Table:MtFits} shows the results of fitting the
$m_{\bot}$ spectra by an exponential $(A_{E}e^{-(m_{\bot}-m)/
T_{E}})$ and Boltzmann $(A_{B}m_{\bot} e^{-(m_{\bot}-m)/T_{B}})$
functions. Both functions give a good representation of the data
but the Boltzmann fit gives a slightly better $\chi ^{2}/dof$. The
inverse slopes of the $\Xi$ particles are the same within
uncertainties and show no apparent increase over the measured
centralities.
 The yields per unit of rapidity are extracted by integrating
each function over the entire $m_{\bot}$ range. The measured $\Xi$
spectra correspond to $\sim$75$\%$ of the total yield and the
$\Omega$ to $\sim$66$\%$. The \xim and \axi yields as function of
$h^{-}$=$dN_{h^{-}}$/d$\eta$\big|$_{|\eta|<0.5}$ appear linear;
such behavior was reported for the $\Lambda$ hyperon at
RHIC~\cite{StarLambda}.

The 10$\%$ most central $m_{\bot}$ spectra are also fit
by a hydrodynamically inspired function ~\cite{Schnederman}.
In this model, all considered  particles are emitted from a
thermal expanding source
with a transverse flow velocity $<\beta_{\bot}>$ at the thermal
freeze-out temperature $T_{\mathrm{fo}}$. As
in~\cite{StarNetProton} a flow velocity profile of
$\beta_{\bot}(r) = \beta_{s}(r/R)^{0.5}$ is used, where $R$ is
the maximum emission radius. The solid lines of
Fig.~\ref{Fig:HydroCont}(a) show the one, two and three sigma
contours for $T_{\mathrm{fo}}$ vs $<\beta_{\bot}>$ for the fit
to the \xim and \axi data combined, with the diamond indicating
the best fit solution
($T_{\mathrm{fo}}=182\pm29$ MeV, $<\beta_{\bot}>=0.42\pm0.06~c$,
$\chi^{2}/dof=13/15$).
 Also shown, as the dashed lines, are the one, two and three sigma
contours for a combined fit to the STAR $\pi$, K, p, and $\Lambda$
data~\cite{StarPion,StarKaon,StarNetProton,StarLambda}, and the
marker is the optimal fit location. Clearly the results for the
two data sets do not overlap indicating that the $\Xi$ baryons,
within this approach, show a different thermal freeze-out
behavior than $\pi$, K, p, and $\Lambda$.
 The current $\Omega$ statistics do not allow to distinguish between
an early decoupling or a common freeze-out with the lighter
species. Fig.~\ref{Fig:HydroCont}(b) shows the mean $p_{\bot}$ for
these particles calculated from the functions which best reproduce
each $m_{\bot}$ spectrum (Bose-Einstein for $\pi$, exponential for
K, hydrodynamically inspired function for p and Boltzmann for
$\Lambda$, $\Xi$ and $\Omega$). The error bars on the experimental
points are statistical and systematic uncertainties added in
quadrature. The band represents the model prediction based on the
three sigma contour for the fit to the STAR $\pi$, K, p, and
$\Lambda$ data while the lower dashed curve shows the prediction
for $T_{\mathrm{fo}}$=$170$ MeV, $<$$\beta_{\bot}$$>$=$0$, {\it
i.e.} a system where thermal and chemical freeze-out coincide and
no transverse flow is developed. From Fig.~\ref{Fig:HydroCont}(a)
it is likely that the $\Xi$ baryons prefer a hotter thermal
freeze-out temperature parameter when compared to that resulting
from fits to the lower mass particles. This is consistent with the
calculated $<p_{\bot}$$>$ of the $\Xi$ and $\Omega$s being below
the solid band in Fig.~\ref{Fig:HydroCont}(b).

If, as indicated by this fit, the $\Xi$ thermal freeze-out occurs
in conjunction with that of chemical freeze-out,
$T_{\mathrm{ch}}\sim$$170$ MeV, this could be an indication that a
significant fraction of the collective transverse flow has already
developed at/before chemical freeze-out, probably at a partonic
stage. Two alternative models have also been proposed for
describing the RHIC spectra. The first~\cite{Broniowski} assumes
that all transverse radial flow is developed at/before
$T_{\mathrm{ch}}$=$T_{\mathrm{fo}}$=165 MeV. The apparent
softening of the lighter mass spectra is then due to contamination
from resonance decay products. A hydrodynamical
approach~\cite{Kolb} is used in the second scenario where the
particle mean free paths are assumed small until thermal freeze-out
around $T_{\mathrm{fo}}$=110 MeV. These large cross-sections
result in even the $\Omega$ developing a sizeable fraction of its
radial flow after hadronization.

\begin{figure}[htb]
\begin{center}
\includegraphics[width=0.5\textwidth]{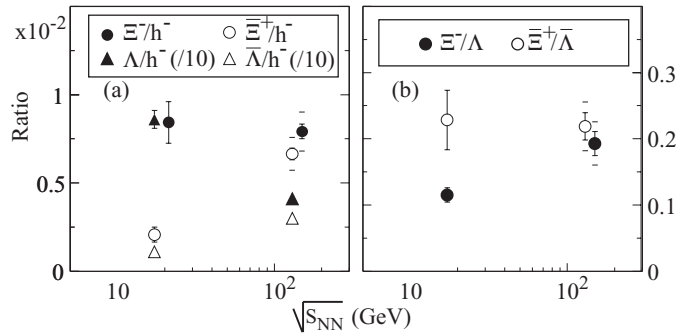}
\caption{ (a) \xim, \axi, $\Lambda$ and $\overline{\Lambda}$ to
$h^{-}$ ratios and (b) \xim/$\Lambda$ and \axi
/$\overline{\Lambda}$ ratios for the most central data as a
function of $\sqrtsNN$.
 The solid lines indicate the statistical uncertainty while the caps
indicate the statistical and systematic uncertainties added in
quadrature.
 Some ratios are slightly shifted along the x-axis for clarity. }
\label{Fig:Ratios}
\end{center}
\end{figure}

 The \axi/$h^{-}$ and $\overline{\Lambda}$/$h^{-}$ ratios for the
most central data, as shown in Fig.~\ref{Fig:Ratios}(a), increase
from SPS energies \cite{WA97Ratio} to RHIC, whereas the \xim/$h^{-}$
ratio stays constant and the $\Lambda$/$h^{-}$ decreases.
 When discussing the baryon ratios  the interplay of increased
strangeness  production and reduction in the net-baryon number,
which was significant at the SPS, has to be considered.
 The proximity of the net-baryon number to zero at RHIC is reflected
in the fact that the ratio of \xim/$h^{-}$ is close to \axi/$h^{-}$.
 The reduction in the net-baryon number has a larger effect on the
$\Lambda$ than on the \xim, as seen in Fig~\ref{Fig:Ratios}(a), and
thus creates the observed rise in the \xim/$\Lambda$ ratio
(Fig.~\ref{Fig:Ratios}(b)).
 It is interesting to note that the \axi/$\overline{\Lambda}$
ratio is a constant from SPS to RHIC indicating that the scale of
the multi-strange enhancement is the same for singly and doubly
strange baryons.

\begin{table}[htb]
\begin{center}
\begin{ruledtabular}
\begin{tabular}{lcccc}
Ratio (x10$^{-3}$) & Data 0-10$\%$ & Stat. & Non-Equil.& HIJING/$B\overline{B}$\\
\hline
\xim $/h^{-}$ & 7.9$\pm$0.4$\pm$1.0 & 7.7& 7.6  & 5.1 \\
\axi $/h^{-}$ & 6.6$\pm$0.3$\pm$0.8 & 6.5 &  6.1& 3.0 \\
$\Omega$ /$h^{-}$ & 2.2$\pm$0.5$\pm$0.4 & 2.9 & 2.8 & 0.29  \\
\xim $/\Lambda$ & 193$\pm$18$\pm$27 & 148 & 190 & 171 \\
\axi $/\bar{\Lambda}$ & 219$\pm$21$\pm$31 &  163 & 207 &  142 \\

\end{tabular}
\end{ruledtabular}
\end{center}
\caption{ Ratios from the 10$\%$ most central data compared to
predictions from a statistical model\cite{ThermMod5} a
non-equilibrium model\cite{Rafelski} and
HIJING/$B\overline{B}$\cite{Hijing}. Uncertainties shown are first
statistical and then systematic.} \label{Table:RatMod}
\end{table}

A fit to the reported 0-10$\%$ centrality particle ratios from
STAR~\cite{StarLambda,StarRatio,StarKaon,StarNetProton,StarPion,StarPhi},
including the multi-strange particle measurements, using the
thermal model described in \cite{ThermMod5} results in a
$\chi^{2}/dof$ of 8.5/9, a $T_{\mathrm{ch}}$ of $181$$\pm$$8$ MeV,
light quark and strange quark chemical potentials of
$11.7$$\pm$$0.6$ MeV and $0.9$$\pm$$1.6$ MeV respectively, and
$\gamma_{s}$=$0.96$$\pm$$0.06$. The fact that $\gamma_{s}$ is
equal to unity, within errors, when used as a free parameter in
the model indicates that a saturation of strangeness production
has been achieved in the most central collisions.
 Table~\ref{Table:RatMod} shows particle ratios for the 0-10$\%$
centrality bin compared to values from three different models: a
fit corresponding to the statistical model used above, another
corresponding to a statistical model that allows for chemical
non-equilibrium  via an over-saturation of both the light quark
and strangeness phase spaces~\cite{Rafelski} and the prediction of
the event generator HIJING/$B\overline{B}$ v1.0 which uses a
gluon-junction mechanism to enhance the transport of baryons to
mid-rapidity~\cite{Hijing}. All ratios are well reproduced by both
statistical models, indicating that the strangeness production is
duplicated well by such approaches. HIJING/$B\overline{B}$,
however, fails to predict the multi-strange to $h^{-}$ ratios, in
particular that of the $\Omega$, which it under-predicts by nearly
an order of magnitude.
 It has previously been shown that HIJING  successfully predicts the
mid-rapidity total charged particle yields~\cite{PhobosNch}
suggesting that the entropy is reasonably well reproduced by this
model.
 The addition  of the gluon-junction mechanism, which was necessary
to replicate the small net baryon yields at
RHIC~\cite{PhenixLambda}, does not sufficiently enhance the
multi-strange baryon yields suggesting that a different physics
mechanism is necessary to model the strangeness production.
 At SPS energies the introduction of final state interactions
helped to account for the observed hyperon enhancement but failed
to reproduce the overall strangeness production ($K^{+}/\pi^{+}$).
The introduction of strong partonic interactions in the initial
state was needed to account for both the hyperon and overall
strangeness production at the SPS~\cite{Vance}.

In summary, the total yield of multi-strange baryons per $h^{-}$ in Au-Au
collisions is increased compared to that at the top SPS energies.
 A chemical analysis of the data indicates that for central
collisions the strangeness phase space is now saturated.
 A fit of the $m_{\bot}$ spectra to a hydrodynamically inspired model
suggests the $\Xi$ baryons thermally freeze out of the rapidly
expanding collision region at a hotter temperature, which is close
to that of chemical freeze-out, and with a smaller transverse flow
than the lighter particle species.
 This suggests that they decouple at an earlier stage of the collision and
thus probe a different dynamical region, but one at which a
sizeable fraction of the transverse flow has already developed.
 All these observations are compatible with the early stage
of the collision being driven by partonic interactions.

We thank the RHIC Operations Group and RCF at BNL, and the NERSC
Center at LBNL for their support. This work was supported in part
by the HENP Divisions of the Office of Science of the U.S. DOE;
the U.S. NSF; the BMBF of Germany; IN2P3, RA, RPL, and EMN of
France; EPSRC of the United Kingdom; FAPESP of Brazil; the Russian
Ministry of Science and Technology; the Ministry of Education and
the NNSFC of China; SFOM of the Czech Republic, DAE, DST, and CSIR
of the Government of India; the Swiss NSF.

\vfill\eject
\end{document}